\documentclass[12pt,epsf,epsfig,psfig]{article}
\usepackage{graphicx}
\usepackage{epsfig}
\usepackage{cite}
\def\U{$\Upsilon$}
\def\u{\Upsilon}

\def\q{q{\bar q}}
\def\Q{Q{\bar Q}}

\def\be{\begin{equation}}
\def\ee{\end{equation}}
\def\be{\begin{equation}}
\def\ee{\end{equation}}

\def\lsim{\raise0.3ex\hbox{$<$\kern-0.75em\raise-1.1ex\hbox{$\sim$}}}
\def\gsim{\raise0.3ex\hbox{$>$\kern-0.75em\raise-1.1ex\hbox{$\sim$}}}

\def\NP{{ Nucl.\ Phys.\ }}
\def\PL{{ Phys.\ Lett.\ }}
\def\PR{{ Phys.\ Rev.\ }}

\def\PRL{{ Phys.\ Rev.\ Lett.\ }}

\def\ZP{{ Z.\ Phys.\ }}
\def\EP{{ Europ.\ Phys.\ J.\ }}

\begin{document}

~~\hfill BI-TP 2010/09

~~\hfill TIFR/TH/10-08

\vskip1cm

\centerline{\Large \bf The QCD Phase Structure at High Baryon Density}

\vskip 1cm

\centerline{\large \bf P.\ Castorina$^a$, R.\ V.\ Gavai$^{b,c}$
and H.\ Satz$^c$}

\bigskip

\centerline{$^a$ Dipartimento di Fisica, Universit{\`a} di Catania,
and INFN Sezione di Catania, Italy}

\medskip

\centerline{$^b$ Tata Institute for Fundamental Research,
Mumbai, India}
        
\medskip
            
\centerline{$^c$ Fakult{\"a}t f{\"u}r Physik, Universit{\"a}t Bielefeld, 
Germany}

\vskip1cm

\centerline{\bf Abstract:}

\bigskip

We consider the possibility that color deconfinement and chiral symmetry 
restoration do not coincide in dense baryonic matter at low temperature. 
As a consequence, a state of massive ``constituent'' quarks would exist 
as an intermediate phase between confined nuclear matter and the plasma of 
deconfined massless quarks and gluons. We discuss the properties of this 
state and its relation to the recently proposed quarkyonic matter.

\vskip1cm

{\bf 1.\ Introduction}

\bigskip

Dense nuclear matter has been a subject of great interest for a long 
time and from quite different points of view. Starting with the medium 
which makes up heavy nuclei, an increase of density led to the regime of 
neutron stars, and once the quark infrastructure of nucleons was introduced, 
to possible quark cores of such stellar states \cite{q-core1,q-core2}. 
Much subsequent work has then addressed the formation of color 
superconducting quark matter \cite{q-core1,super}. In the present paper, 
we want to return to the idea that
besides these states, at high baryon density a plasma of massive deconfined 
quarks can still exist as yet another form of strongly interacting 
matter \cite{Shuryak,Pisarski,Baym-Bie,Cley,Schulz,Drago,Toki,Guo}. 

\medskip  

The essential features of hadrons are color confinement and
spontaneous chiral symmetry breaking. The former binds colored quarks
interacting through colored gluons to color-neutral hadrons. The latter
brings in pions as Goldstone bosons and gives the essentially massless quarks 
in the QCD Lagrangian a dynamically generated effective mass. Both features 
will come to an end in hadronic matter at
sufficiently high temperatures and/or baryon 
densities, though not {\sl a priori} simultaneously. However, rather 
basic arguments suggest that chiral symmetry restoration occurs either 
together with or after color deconfinement \cite{B-C}.

\medskip

\begin{figure}[htb]
\centerline{\psfig{file=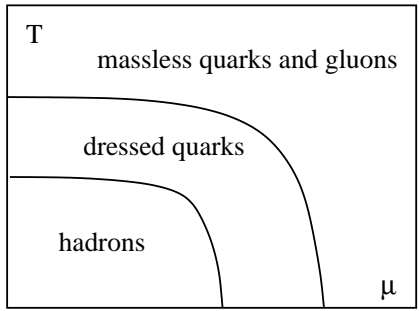,width=5.5cm}}
\caption{Three-state scenario of QCD matter \cite{Baym-Bie}}
\label{three-state}
\end{figure}

It thus appears conceivable that QCD could lead to a three-state phase
structure as function of the temperature $T$ and the baryochemical
potential $\mu$, as shown in Fig.\ \ref{three-state} \cite{Baym-Bie}. 
In such a scenario,
color deconfinement would result in a plasma of massive ``dressed''
quarks; the only role of gluons in this state would be to dynamically
generate the effective quark mass, maintaining spontaneous chiral 
symmetry breaking. At still higher $T$ and/or $\mu$, this gluonic 
dressing of the quarks would then ``evaporate'' or ``melt'', leading to 
a plasma of 
deconfined massless quarks and gluons: the conventional QGP, with 
restored chiral symmetry.
Evidently, this view of things ignores the possibility of bosonic
diquark binding and condensation as well as that of the color 
superconductivity states which could result as a consequence; 
we shall return to these aspects later on. 

\medskip

\medskip

As already noted, the possible existence of a phase of quarks with a 
dynamically generated effective mass has often been considered 
before \cite{Shuryak,Pisarski,Baym-Bie,Cley,Schulz,Drago,Toki,Guo}. 
However, in recent years finite temperature lattice QCD investigations 
have provided an impressive amount of further information which can be 
used in determining the properties of constituent quarks 
\cite{K-Z,Doering,Huebner}. In addition, the proposal for a state 
of ``quarkyonic'' matter in the limit of a large number of color degrees 
of freedom \cite{quarky1,quarky2} has also renewed the interest
in a possible constituent quark phase.

\medskip

The basic idea in our considerations will be that a medium of constituents
endowed with an intrinsic spatial scale $L$ can only exist as long as its 
density remains below $1/L^3$. In strong interaction physics, this had
first led to the prediction of an upper density limit for mesonic matter 
\cite{Pomeranchuk}. In the past decades it has found a quantitative formal 
basis in percolation theory \cite{Stauffer}. The natural starting point is 
thus the determination of the intrinsic scales for the given medium. 

\medskip

Hadronic interactions lead to two intrinsic scales, one on the hadronic and 
the other on the quark level. Both can be expressed in different ways, and
we shall elaborate on this. However, since 
we want to resort to geometric arguments, size parameters are the most
useful. One scale then is given by the confinement radius $R_h\simeq 1$ fm, 
defining the range of the strong force and thus also the size of hadrons. 
As long as the density $n$ of the medium remains below
$n_c^h \simeq  1/V_h^3$, with $V_h=4\pi~\!R_h^3/3$, 
it is expected to be of hadronic nature. Inside a hadron, the valence quark 
constituents acquire a dynamically generated effective mass $M_q$ and size 
$R_q$. We shall show in the next section that both theoretical and
experimental studies indicate $M_q \simeq 0.3-0.4$ GeV and $R_q \simeq 0.3$ 
fm. In the density region $n_c^h < n < n_c^q \simeq 1/V_q^3$, 
with $V_q=4\pi ~\!R_q^3/3$, we then
expect a plasma of deconfined massive quarks, while for $n>n_c^q$,
the medium will become the conventional quark-gluon plasma, with pointlike 
massless quarks and gluons as constituents.  If both the parameters $R_h$ 
and $R_q$ were independent of temperature and baryon density,
the critical points in the phase diagram of Fig.\ \ref{three-state} could 
thus be easily defined in terms of corresponding percolation points. 

\medskip

However, it is well-known that the latter premise is not generally correct.
Lattice 
studies at or near $\mu\!=\!0$ have shown that here color deconfinement and 
chiral symmetry restoration coincide \cite{dec=ch}. Hence in a medium 
of vanishing baryon density, the mass of the constituent quark 
vanishes at the deconfinement point $T_c$. This is in accord
with the formation of the quark dressing through a polarization cloud
in the surrounding gluonic medium; as the temperature approaches
$T_c$, the screening radius defining cloud size decreases rapidly, and 
at $T_c$, the cloud has ``evaporated'', leaving pointlike quarks and gluons. 
There thus remains only one scale, which defines the simultaneous onset of
color deconfinement and chiral symmetry restoration.
Since the physical reason for this is apparently the 
presence of a hot gluonic environment, there is little 
reason to expect a similar behavior at low $T$ and large $\mu$. The 
aim of our study is therefore to introduce in this region an intermediate 
plasma of massive quarks, separating hadronic matter and QGP by a state 
of quark deconfinement and broken chiral symmetry.

\medskip  

We begin by considering nature and properties of constituent quarks, 
followed by a discussion of deconfinement and chiral symmetry
restoration as consequence of the density of the relevant constituents.
With these results as input, we then study the possible structure of 
a massive quark phase, and finally compare it to that of quarkyonic 
matter. 
 
\bigskip

{\bf 2.\ Constituent Quarks}

\bigskip

There are two different regimes for the quark infrastructure of
hadrons, depending how we probe. Relatively hard probes, such as
deep inelastic lepton-hadron scattering or hadron-hadron interactions
at large momentum transfer, lead to massless pointlike quarks and gluons.
In this regime, the parton model with hadronic quark and gluon
distribution functions provides a suitable description. On the other
hand, soft interactions, as seen in minimum bias proton-proton or pion-proton
interactions, suggest that mesons/nucleons consist of two/three
``constituent'' quarks  having a a size of about 0.3 fm and a mass of 
about 0.3 - 0.4 GeV. Here many features are well accounted for by the additive 
quark model \cite{additive}. We can thus imagine that inside a hadron, a 
quark polarizes the gluon medium in which it is held through color 
confinement, and the resulting gluon cloud forms the constituent quark 
mass $M_q$ \cite{Novi,S-V}. 
 
\medskip

This picture is today found to be quite compatible with heavy quark 
correlation studies in finite temperature lattice QCD at vanishing
baryon density. By evaluating Polyakov loop 
correlations in a QCD medium of two or three light quark flavors
below deconfinement ($T < T_c$), one obtains the free energy $F(r,T)$
as function of the quark separation distance $r$. In the low 
temperature limit, $F(r,T=0)$ saturates beyond a separation of 
$r \simeq 1.5$ fm, converging to the value
$F(\infty,T=0) \simeq 1.2 \pm 0.1$ GeV
\cite{K-Z}. This result is quite universal; it is obtained by separating
a heavy quark-antiquark pair, where the separation requires the 
formation of a light $\q$ pair to assure color neutrality. It is obtained 
as well even if we separate a heavy quark-quark pair, where the formation
of a light antiquark-antiquark pair is necessary \cite{Doering}. Moreover, 
it is reached
for any color channel (singlet or octet for $\Q$, antitriplet or 
sextet for $QQ$). The large $r$ behavior of all cases coincides; any
uncertainty in the numerical value of $F(\infty,T=0)$ is due to the
necessity to extrapolate to $T=0$ and to uncertainties in the 
normalization.

\medskip

To create an isolated heavy-light quark system, we thus need a
gluonic energy input $F_g$, with 
\be
F(\infty,T=0)  = 2~\!F_g \simeq 1.2~{\rm GeV}.  
\ee
There are various ways to cross-check this result. If we consider a QCD
medium in which light quarks are not so light, having a specified 
bare mass $m_q$, then the result should be
\be
F(\infty,T=0)  = 2~\!F_g + 2~\!m_q.  
\ee
For a range of different values of $m_q$, one always obtains
$F_g \simeq 0.6$ GeV \cite{KLP}. 
Lattice studies have also considered
the quark separation of a nucleon, bringing all three quarks far
apart \cite{Huebner}. The resulting free energy now becomes
\be
F_3(\infty,T=0)  = 3~\!F_g,  
\ee
and the lattice results again lead to the same value of $F_g$
as found above. 

\medskip

Experimentally, a similar check can be performed, comparing 
the mass values of the open charm and open bottom mesons
to those of the charm and bottom quarks. This yields
\be
M_D - m_c = F_g = 0.60 \pm  0.10~{\rm GeV}
$$$$
M_B - m_b = F_g = 0.53 \pm 0.15~{\rm GeV},
\ee
using the relevant mass values as given by the PDG listing \cite{PDG}, 
and thus confirms that the resulting value does not depend on the
quark source mass. The sizes of the heavy-light mesons are expected to 
be of hadronic scale, so that a gluonic polarization system of spatial
extent makes sense. The spatial extent of quarkonia, however, decrease 
inversely proportional to the heavy quark mass. Hence the \U~presents 
to the gluonic medium effectively a color singlet system of negligible 
extent, and it should therefore produce no polarization. In accord with 
this, one has
\be
M_{\u} - 2~ m_b = 0.06 \pm 0.31~{\rm GeV},
\ee
so that \U~the mass is essentially twice the bare bottom quark mass.

\medskip

The same value persists for light-light quark systems, giving the 
correct vector meson masses (for both non-strange and strange
ground states) as well as the correct different ground state baryon 
masses. In the case of hadrons containing strange quarks, the 
relevant number of strange quark masses ($m_s\simeq 100$ MeV)
has to be added.

\medskip

What is the meaning of $F_g$? One possible and rather widely accepted 
interpretation is that it is the ``mass'' or the energy content 
of the gluonic string connecting quark and antiquark. With
\be
F_g  \simeq \sigma~\!R_h \simeq 0.6 - 0.8~{\rm GeV}
\ee
and using $\sqrt \sigma=0.4$ GeV and $R_h = 0.8 - 1.0$ fm, this
does lead to the correct value of $F_g$, at least in the case of
mesons; baryons are not so easily dealt with. 

\medskip

We want to consider here instead a scenario in which $F_g$ is the sum of 
the gluonic dressing masses of two constituent quarks. Then both mesons and 
baryons can be treated on equal footing,
giving us
\be
M_q = {F_g\over 2} + m_q \simeq 0.3 - 0.4~{\rm GeV},
\ee 
where $m_q$ denotes the bare quark mass; the last term thus
corresponds to the light quark limit. We emphasize that the
constituent quarks retain their intrinsic quantum numbers; the
gluon cloud thus is color-neutral and without any spin.  

\medskip

Such an interpretation is, as already mentioned, supported by
the additive quark model \cite{additive}. In a collision energy 
range of about $\sqrt s \simeq 5-20$ GeV, in which hard processes 
do not yet play a significant role, the total cross sections for 
proton-proton and pion-proton collisions are given as
\be
\sigma_{pp}=3\times 3 \sigma_{qq} \simeq 38~{\rm mb},~
,~
\sigma_{\pi p}=2\times 3 \sigma_{qq} \simeq 25~{\rm mb}.
\label{addi}
\ee
The predicted ratio 3/2 between pion and proton projectiles is
seen to be in accord with the data; moreover, $\sigma_{pp} = 
\pi~\!R_h^2$ leads to $R_h \simeq 0.9$ fm for the hadronic radius. 
Using eq.\ (\ref{addi}), we obtain
\be
 \sigma_{qq} \simeq 3.3~{\rm mb} ~\to~R_q \simeq 0.33~{\rm fm}
\ee
for the corresponding constituent quark sizes in the case of
light bare quarks; we return to the more general case of $m_q \gg 0$
shortly. A similar constituent quark radius was also obtained through
partonic arguments \cite{S-V}.

\medskip

\begin{figure}[h]
\centerline{\psfig{file=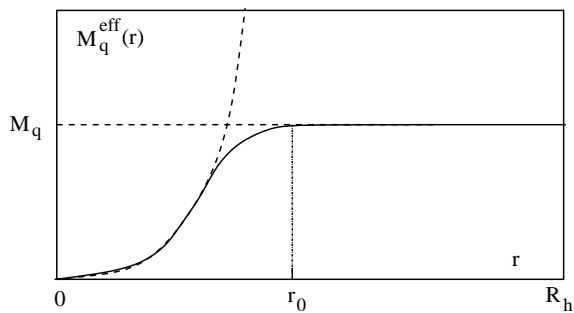,width=7.5cm}}
\caption{Effective quark mass $M_q^{\rm eff}(r)$ as seen from a distance 
$r$ \cite{Shuryak}.}
\label{c-mass-r}
\end{figure}

As mentioned, we consider the constituent quark to be made up of
the bare quark and the gluonic polarization cloud surrounding it.
This means that as we move a distance $r$ away from the pointlike quark,
we find an effective quark mass $M_q^{\rm eff}(r)$, depending on how
much of the cloud we include at a given $r$. Screening in the non-abelian 
gluon medium limits the size of the cloud, so that beyond $r_0 \simeq 0.3$ fm,
the cloud mass saturates, with the constituent quark mass $M_q$ as 
saturation limit. The resulting behavior \cite{Shuryak} is illustrated 
in Fig.\ \ref{c-mass-r}, with $R_h \simeq 1$ fm denoting the 
radius of color confinement. 

\medskip

\medskip

The conceptual scenario just discussed is supported by perturbative QCD 
estimates \cite{Politzer}. In the chiral limit ($m_q \to 0$), 
the effective quark mass $M^{\rm eff}_q(r)$ at scale $r$ is determined 
by the (non-perturbative)
chiral condensate $\langle{\bar\psi}\psi\rangle$ and renormalization factors,
\be
M_q^{\rm eff}(r) = 4~g^2(r)~r^2 \left[{g^2(r) \over g^2(r_0)}\right]^{-d}
\langle{\bar\psi}\psi(r_0) \rangle,
\label{m1}
\ee
where $r_0$ denotes a reference point for the determination of 
$\langle{\bar\psi}\psi(r_0) \rangle$; it is something like the meeting 
point of perturbative and non-perturbative regimes. For the case of 
massless quarks with $N_c=3,~N_f=3$, the anomalous dimension is 
$d=4/9$, and
\be
g^2(r) = {16 \pi^2 \over 9} {1 \over \ln[1/(r^2 \Lambda_{\rm QCD}^2)]}
\label{coupling}
\ee
is the running coupling at scale $r$. The constituent quark mass $M_q$ 
is defined as the solution of eq.\ (\ref{m1}) at the scale $r=1/2M_q$
\cite{Politzer}. This allows us to rewrite eq.\ (\ref{m1}) in the form
\be
M_q^{\rm eff}(r) = M_q \left[ 4~\!M_q^2~\!r^2 \right] 
\left[{g^2(1/2M_q) \over g^2(r)}\right]^{d-1},
\label{m2}
\ee
showing how the effective quark mass decreases at short 
distances $r < 1/2M_q$, starting from the constituent quark value.
The value itself is determined by the non-perturbative
chiral condensate at some reference point $r_0$. From
eq.\ (\ref{m2}) we have
\be
{M_q^3 \over \langle{\bar\psi}\psi(r_0) \rangle}=\left({16 \pi^2 \over 9}
\right)\left(\ln [1/(r_0^2 \Lambda_{QCD}^2)]\right)^{-4/9}
\left(\ln[4M_q^2/ \Lambda_{QCD}^2]\right)^{-5/9}.
\label{m3}
\ee
As reference scale we use $r_0= 1/2M_q$, since that is where the 
perturbative evolution stops and the non-perturbative regime starts.
We thus obtain
\be
{ M_q^3 \over \langle {\bar\psi}\psi(r_0) \rangle}=
{16 \pi^2 \over 9}~ {1 \over \ln (4M_q^2 / \Lambda_{QCD}^2)},
\label{m4}
\ee
showing that the constituent quark mass indeed provides a scale parameter
for chiral symmetry breaking.
To get an estimate for its value, we use 
$\Lambda_{QCD} = 0.2$ GeV and $\langle {\bar\psi}\psi(r_0) \rangle^{1/3} 
= 0.2$ GeV; this yields as solution of eq.\ (\ref{m4})
$M_q = 375$ MeV; the corresponding constituent quark radius
becomes $R_q = r_0 = 0.26$ fm. At the point $r_0=1/2M_q$, eq.\ 
(\ref{coupling}) gives for the strong coupling
\be
\alpha_s(r_0) = {g^2(r_0) \over 4 \pi} \simeq 0.5 ;
\label{alpha}
\ee
we have assumed a perturbative evolution in $r$ up to this point, which
may be subject to some doubt. The value (\ref{alpha}) is in approximate 
agreement with the value obtained in static quark lattice studies 
\cite{KKPZ}, which indicate, however, remaining non-perturbative 
contributions at this $r$. Nevertheless, it seems remarkable that 
the results agree rather well with the estimates obtained above
from lattice calculations as well as from experiment. 

\medskip

The constituent quark radius is thus found to be $R_q = 1/2~\!M_q$
in the chiral limit. More generally, we then expect
\be
R_q \simeq {1 \over 2(M_q + m_q)},
\ee
leading to a decrease in size with increasing bare mass. This is
in accord with the decreasing cross sections for the interaction of
strange or charm mesons \cite{Povh}.    

\medskip

The effective constituent quark mass is thus determined by the 
size and energy density of the gluon cloud, or equivalently,
by the chiral condensate value in the non-perturbative region.
How do these quantities change with temperature in a hadronic medium 
at vanishing baryon density?
 
\medskip

\begin{figure}[h]
\centerline{\psfig{file=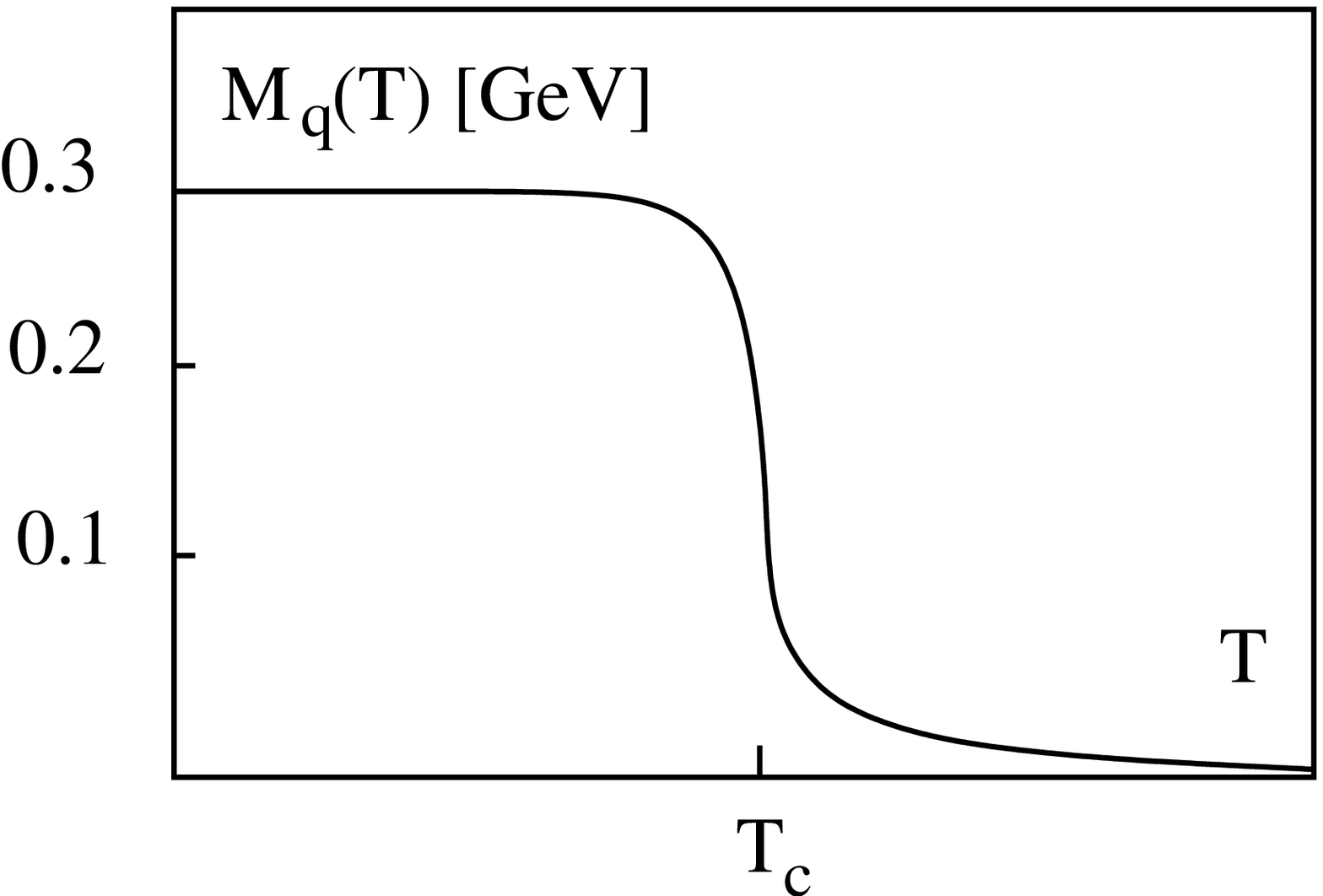,width=6cm}\hskip1cm
\psfig{file=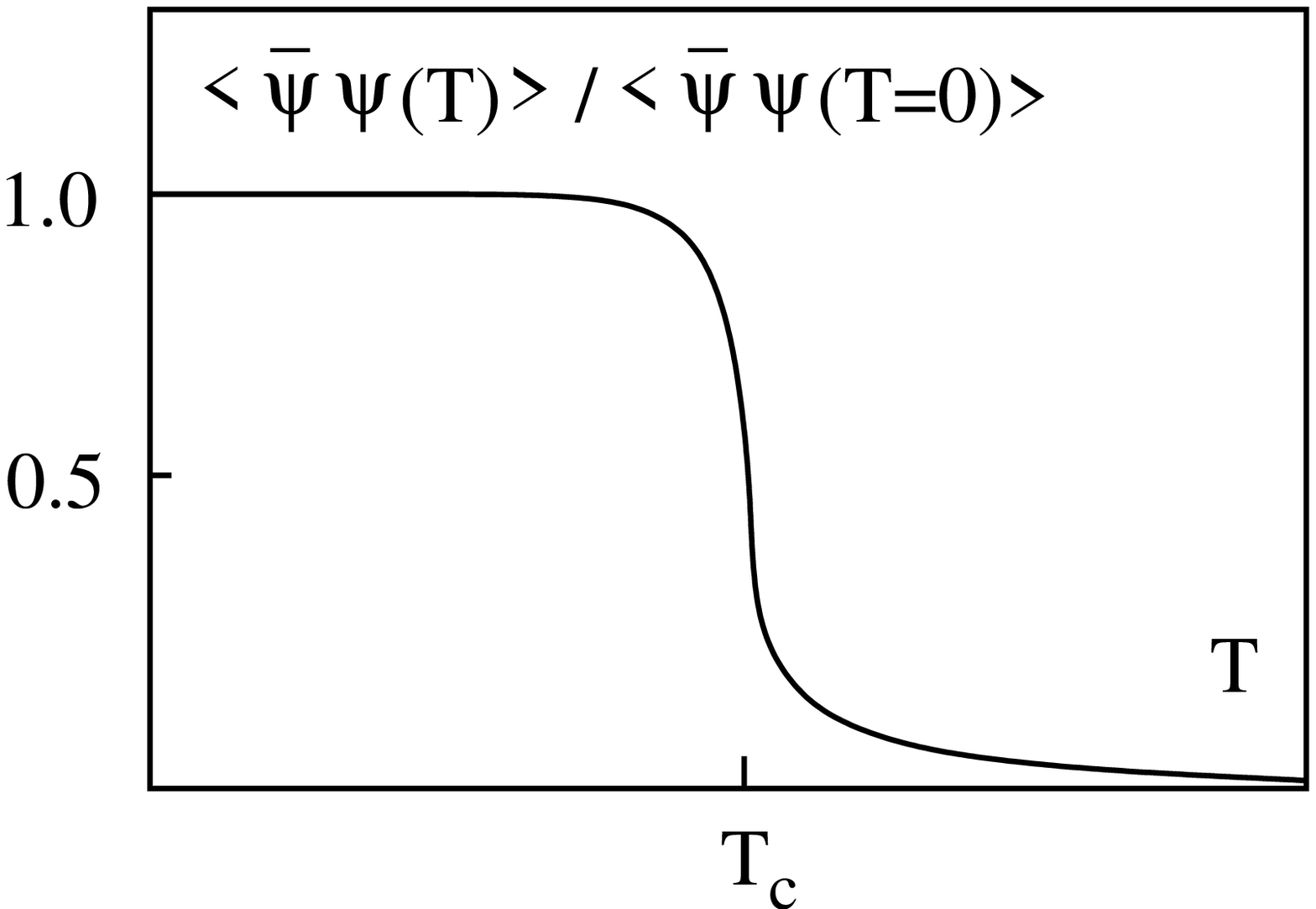,width=6cm}}
\caption{Constituent quark mass $M_q(T)$ (left)
and chiral condensate $\langle {\bar\psi}\psi(T) \rangle$ (right)
as function of temperature $T$.}
\label{Temp-dep}
\end{figure}

This can again be deduced from heavy quark correlation studies as 
function of the temperature of the medium. They show that the effective 
mass of the gluon cloud of an isolated static color charge (obtained by 
separating a static $\Q$ pair), starting from confinement values around
300 MeV, drops sharply at $T\simeq T_c$ \cite{dec=ch}. This is accompanied 
by a corresponding drop of the screening radius. We thus expect the 
effective quark mass to show the temperature dependence illustrated 
in Fig.\ \ref{Temp-dep}.
  
\medskip

\medskip

Complementary to this, the temperature dependence of the chiral condensate
is determined directly in finite temperature lattice QCD. Its behavior is 
also shown in Fig.\ \ref{Temp-dep}; it is seen that at the deconfinement point,
the chiral condensate vanishes as well \cite{dec=ch}. This is in accord 
with the idea that at this point, the gluon cloud around the give quark 
has essentially evaporated.

\medskip

These considerations show that there are two distinct ways to reach
chiral symmetry restoration. On the one hand, even for an interquark distance 
$2~\!R_h$ well above $2~\!R_q$, a sufficiently hot medium will through 
gluon screening cause the effective quark mass to vanish, as shown in 
Fig.\ \ref{Temp-dep}. On the other hand,
when a cold medium becomes so dense that the average interquark distance is 
$2~\!R_q$ or less, the quarks form a connected cluster containing pointlike
bare quarks.

\medskip

We note in passing that the two scales, $R_h$ and $R_q$, have also been
considered as the quark and gluon confinement scales, respectively. This
implies that color-neutral hadrons have size $R_h$, whereas color-neutral
glueballs have the much smaller intrinsic size $R_q$, and the spatial
ground state glueball size is indeed in most calculations found to   
be about $R_q \simeq 0.3$ fm.

\medskip

We want to argue in the following section that at $\mu=0$, color 
deconfinement sets in at relatively large interquark separation, 
but in a hot gluonic environment. As a consequence, the gluon cloud 
giving rise to the effective quark mass has evaporated, causing 
deconfinement and chiral symmetry restoration to coincide. 
For the other extreme, for $T \simeq 0$, at the color deconfinement 
point the interquark distance is also still well above $R_q$.
But now there is no hot gluonic medium to melt the gluon cloud, 
so that the cold deconfined medium will be a plasma of massive quarks.
Only when its density is increased much more, to the percolation
point of the massive constituent quarks, will the medium effectively 
consist of massless pointlike quarks.

\bigskip

{\bf 3.\ Percolation and Phase Boundaries} 

\bigskip

\begin{figure}[h]
\centerline{\psfig{file=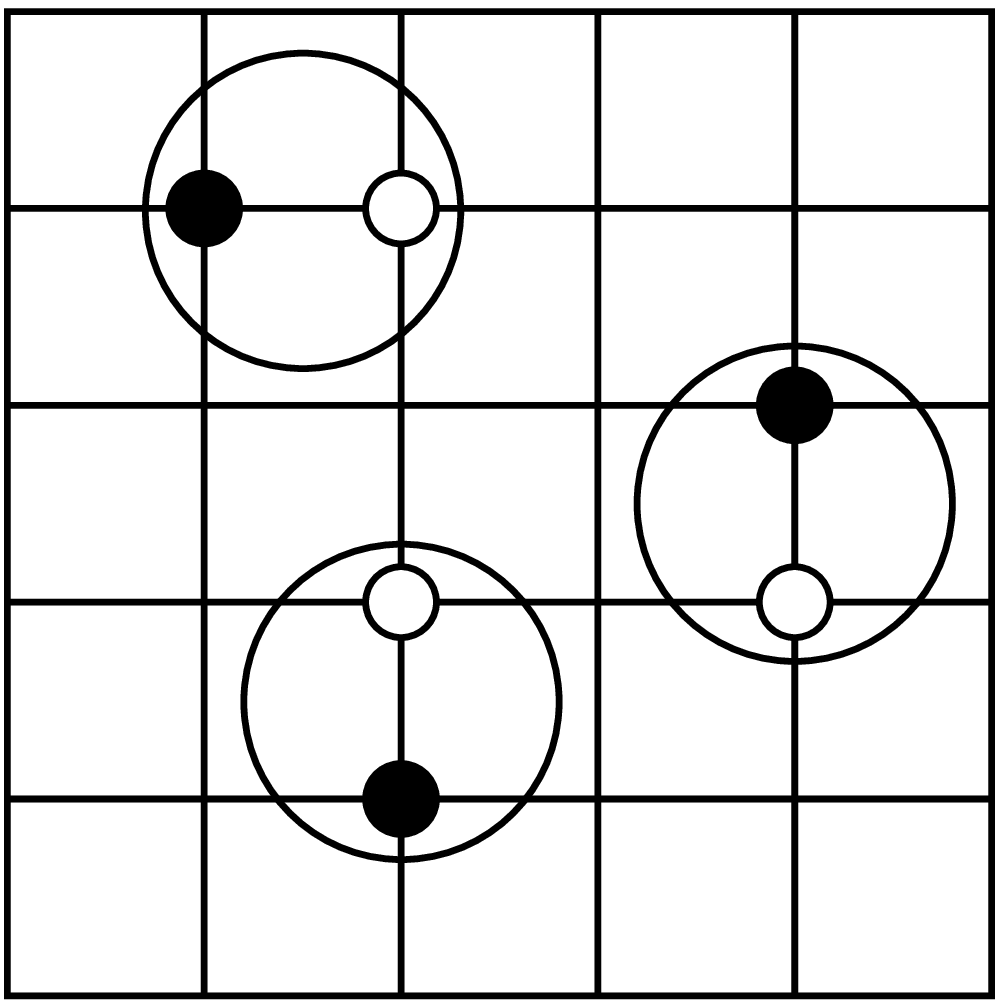,width=3cm}\hskip3cm
\psfig{file=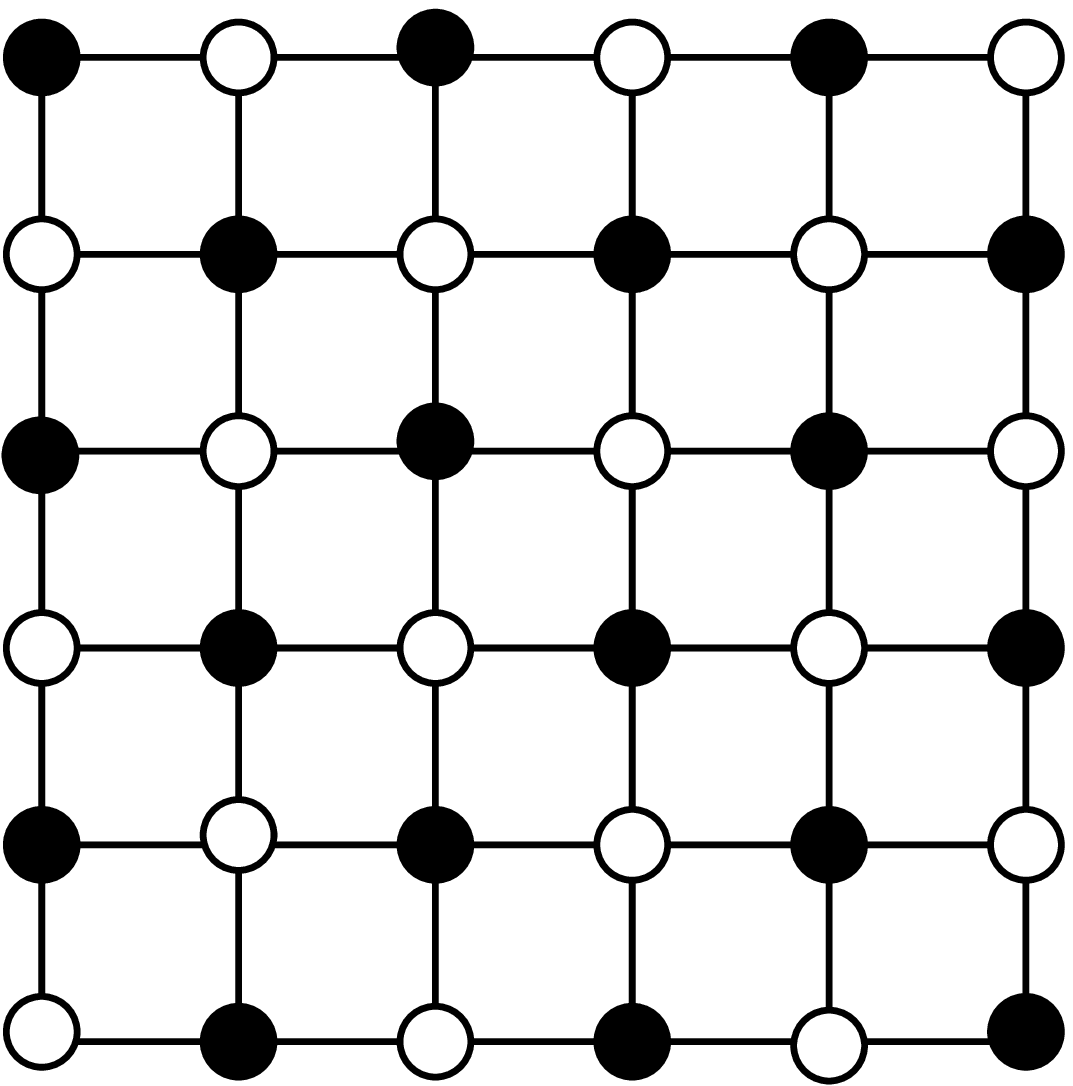,width=3cm}}
\caption{Schematic view of the transition from confined (left) to
deconfined matter (right)}
\label{dense}
\end{figure}

Hadrons are color-singlet quark-antiquark or three-quark states having
a characteristic spatial extension of about one fermi. In a hadronic medium,
the quark constituents are restricted to the corresponding volume by
color confinement, making it impossible for a given quark to separate 
more than about a fermi from its partner(s) in a color-singlet. When the 
density of a gas of hadrons is increased sufficiently, by raising either 
the temperature or the baryon density, a quark constituent in a given 
hadron will eventually find quarks and antiquarks from other hadrons as 
close by as its original partner(s). At this point, one can no longer 
define specific hadrons and hence the concept of hadronic matter becomes 
meaningless; the hadron gas has become a plasma of deconfined colored 
quarks and antiquarks.  This is illustrated schematically in Fig.\ 
\ref{dense} for a discretized world. As mentioned, such a density limit 
to hadronic matter was first suggested by Pomeranchuk, well before the 
advent of the quark model of hadrons \cite{Pomeranchuk}.

\medskip

\medskip

In the chiral limit, both mesons and baryons have an intrinsic spatial
size of about 1 fm. Hence in both cases, the formation of percolating 
clusters provides a natural limit to the hadronic form of strongly 
interacting matter \cite{Baym,Celik,Magas,CRS}. We note that for
spin dynamics, the resulting percolation theory can be rigorously 
formulated \cite{FK,CK}; for QCD, however, there were only some first
approximative attempts \cite{Fortunato}, and a definitive theory
is still lacking. In particular, both the definition of clusters
(using bond weights) and the relevant thermal distribution law
(with more than next neighbors) are not yet specified.
Our argumentation thus has to remain on a qualitative level.

\medskip

For mesons, increasing the density eventually leads to a medium in
which the distance between a quark constituent from one hadron and 
an antiquark from another is equal to or less than the typical hadronic 
size, so that defining specific quark-antiquark pairs as hadrons ceases 
to make any sense. The percolating medium becomes a plasma
of deconfined quarks, antiquarks and gluons. In the baryon-rich region,
the increase of density of hard-core nucleons leads to ``jamming'', i.e., 
a restriction in the mobility of the nucleons 
\cite{KS,jamming}. But here as well we reach eventually the formation of 
a percolating medium \cite{Kratky}, in which the overall quark density
is too high to define individual nucleons. We thus consider the limit
of confinement in the $T-\mu$ diagram of strongly interacting matter
to be universally defined by percolation of the relevant hadron species 
in the given region \cite{CRS}. It should be emphasized that such a 
percolation limit 
is a well-defined geometric form of critical behavior, specified by
critical exponents and leading to universality classes, just as found 
for thermal critical behavior \cite{Stauffer}. The essential difference 
is that geometric singularities (formation of infinite clusters) do not
necessarily imply non-analytic behavior for the partition function.  

\medskip

At $\mu=0$,
the percolation density for permeable hadrons (mesons and low density
baryons) of size $V=4\pi R_h^3/3$ is found to be \cite{CRS}
\be
n_M \simeq {1.2 \over V_h} \simeq 0.6~{\rm fm}^{-3},
\label{perc-dens-m}
\ee
with $R_h \simeq 0.8$ fm for the hadron radius.
This is the density of the percolating cluster at the onset of
percolation, or, equivalently, the density at which the medium 
no longer allows a spanning vacuum. 
The corresponding temperature can be obtained once the hadronic 
medium is specified. For $\mu \simeq 0$, this medium shows abundant
resonance formation, and such interactions can be taken into account
\cite{Beth,DMB} by replacing the interacting medium of pions and nucleons by
an ideal gas of all observed resonance states. For such an ideal gas 
one finds as deconfinement temperature \cite{CRS}
\be
T_{\rm dec} \simeq 180 ~{\rm MeV},
\ee
which agrees well with the value presently obtained in finite
temperature lattice QCD. 

\medskip

The resulting medium at this point is, however,
strongly non-perturbative. In fact, if we consider as above 
(see eq.\ (\ref{alpha})) the onset of perturbative behavior to be 
given by $\alpha_s \simeq 1/2$, the non-perturbative regime extends
up to $T \simeq 4~T_c$. This is again in accord 
with lattice results, showing that above $3-4 ~T_c$ the interaction
measure $(\epsilon - 3~\!P)/T^4$ reaches perturbative behavior 
\cite{inter-pert}.

\medskip

Assuming mesons to be the dominant constituents, the density 
(\ref{perc-dens-m}) implies for the average separation between quarks
and/or antiquarks at the deconfinement point
\be
d_q^M \simeq {1 \over n_M^{1/3}} \simeq 1.2 ~{\rm fm},
\ee
as illustrated in Fig.\ \ref{dense}. 

\medskip

At the other extreme, for $T=0$, we have to consider the percolation of
nucleons with a hard core \cite{Kratky,CRS}. Assuming a hard core radius
$R_{\rm hc} = R_h/2$, one obtains for the critical density
\be
n_B \simeq {2 \over V_h} \simeq 0.9~{\rm fm}^{-3},
\label{perc-dens-n}
\ee
a value about 30\% higher than that for permeable hadrons, as consequence
of the baryon repulsion. This value can be used to obtain the percolation 
value of the baryochemical potential, using a van-der-Waals approach
to account for the repulsion in the determination of the density as
function of $\mu$ \cite{CRS}. As result, one obtains
\be
\mu_{\rm dec} \simeq 1.1~{\rm GeV}
\ee
as deconfinement point for $\mu$. The separation between quarks at this 
point becomes
\be
d_q^B \simeq {1 \over n_B^{1/3}} \simeq 1.0~{\rm fm},
\ee
slightly less than at $T=0$, due to the higher density. We note that
since $\mu \geq M$, where $M$ denotes the nucleon mass, this leaves
as function of $\mu$ a rather small window 
\be
M \leq \mu \leq 1.2~M 
\ee
for the range of confined baryonic matter at T=0. This window contains
essentially all strongly interacting 
matter in the real world, from nuclei
to neutron stars. The corresponding density range runs from
$n_0 \simeq 0.17$ fm$^{-3}$ as standard nuclear matter density
to the deconfinement value (\ref{perc-dens-n}) of about 5~$n_0$.
 
\medskip

We thus confirm through percolation arguments that color deconfinement 
is expected to set in at hadron densities for which in general the quark 
constituents are separated by about 1 fm. At this density, any partitioning
into hadrons becomes meaningless, and we have a medium of deconfined
quarks of mass $M_q \simeq 0.4$ GeV and size
$r_0 \simeq 0.3$ fm, separated by a distance $r\simeq 1~{\rm fm} >
r_0$. Hence in the density range corresponding to
$r_0 \leq r \leq 1$ fm, the quarks can retain their effective
constituent mass, so that the deconfined medium now is a plasma of
quarks of finite mass and spatial extent, with continued chiral 
symmetry breaking. A sufficient further increase in density will 
eventually lead to overlap and percolation of the constituent
quarks. We assume that beyond this percolation point, chiral symmetry
is effectively restored. Let us see what density the above obtained 
value of $R_q$ leads to.

\medskip

\begin{figure}[htb]
\centerline{\psfig{file=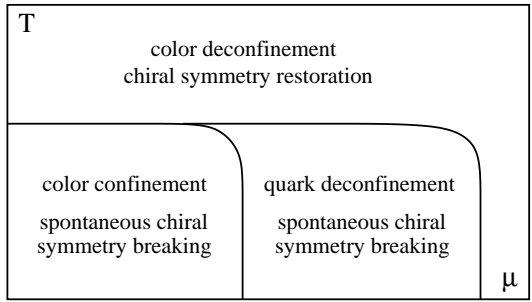,width=7cm}}
\caption{Phase structure of strongly interacting matter}
\label{newphase}
\end{figure}

At $T=0$, the density for a system of quarks of mass $M_q$
is given by
\be
n_q(\mu_q) = {2 \over \pi^2}(\mu_q^2 - M_q^2)^{3/2},
\label{q-dens}
\ee
with $\mu_q = \mu/3$ for the quark chemical potential.
The percolation condition for quarks of radius $R_q$
(see eq.\ (\ref{perc-dens-m})) 
\be 
n_q^{\rm ch} = {1.2 \over (4\pi~\!R_q^3/3)} \simeq {0.29 \over R_q^3}
\ee
then defines the onset of chiral symmetry restoration. 
With $R_q \simeq 0.3$ fm, we obtain
\be
n_B^{\rm ch} \simeq 3.5~\!{\rm fm}^{-3} \simeq 3.9~\!n_B^{\rm dec},
\ee
indicating that the baryon density threshold for chiral symmetry 
restoration is about four times higher than that for color 
deconfinement. The corresponding value for the baryochemical 
potential is found to be $\mu_B^{\rm ch} \simeq 2.2$ GeV, to
be compared to $\mu_B^{\rm dec} \simeq 1.1$ GeV. Using the 
$\mu$ counterpart of the two-loop form (\ref{coupling}), the
strong coupling $\alpha_s$  has now dropped to the value 
$\alpha_s(\mu_c^{\rm ch}) \simeq 0.5$. The resulting
phase structure is schematically illustrated in Fig.\ \ref{newphase}.

\medskip

\medskip

\begin{figure}[htb]
\centerline{\psfig{file=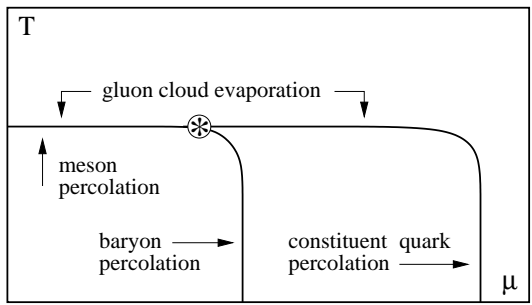,width=7cm}}
\caption{Limits of strongly interacting matter}
\label{percolimit}
\end{figure}

We close this section with some comments on the origin of the limiting
curves. The hadronic phase ends both in $T$ and in $\mu$ by percolation,
for permeable mesons and hard-core nucleons, respectively. The
temperature limit coincides with the gluon evaporation limit, at which
the constituent quark mass vanishes in a non-perturbative way. This
presumably will also occur in the phase of massive quark constituents,
so that the temperature limit of that phase is defined by gluon
evaporation. The low temperature density limit there, however, is 
determined by constituent quark percolation. The behavior in the different 
regions is summarized in Fig.\ \ref{percolimit}. 

\medskip

\medskip

The nature of the limits for the hadronic phase, as given in a percolation
picture, was elaborated in \cite{CRS}. The mesonic percolation limit is
in general purely geometric and not a thermal phase transition, although
given specific dynamics, it can be \cite{FK,CK}. The form of the restoration 
of chiral symmetry there depends on $N_f$. The percolation of hard-core
nucleons in the presence of an overall nucleon attraction leads to a
Van der Waals situation and thus to a first order phase transition \cite{CRS},
ending in a second order point $T^*,\mu^*$. In view of the 
adjacent constituent quark phase this can become a triple point,
as illustrated in Fig.\ \ref{percolimit} \cite{quarky1,quarky2}.  

\bigskip

{\bf 4.\ The Structure of a Plasma of Massive Quarks} 

\bigskip 

Our considerations thus suggest the existence of a plasma of massive
deconfined quarks between the hadronic matter state and the quark-gluon
plasma. In this state, quarks are deconfined; gluons, however, 
are ``bound'' into the constituent quark mass and thus remain
in a sense confined. The quark dressing is made up of gluons which
form a color-neutral cloud, so that the
massive quarks retain their fundamental color state as well as their
other intrinsic quantum numbers. The effective
degrees of freedom in the resulting quark plasma thus are just those
of massive quarks. Its
lower limit in baryon density is defined by the onset of
vacuum formation, forcing the quarks to bind into color neutral nucleons.
The corresponding high density limit is given by the percolation point 
of the (spatially extended) quarks, beyond which there is a connected
medium containing bare quarks and gluons. Finally, increasing the
temperature at fixed $\mu$ leads here, just as in the hadronic phase,
to an evaporation of the gluonic dressing of the quarks and thus to
a restoration of chiral symmetry. 

\medskip

The constituent quarks in the deconfined medium
will in general be interacting with each other. Of particular interest
here is the presence of a $qq$ attraction, which at sufficiently 
low temperatures could lead to the formation of bound colored bosonic 
$qq$ states (``diquarks''). Baryons have in fact often been considered
in terms of two quarks bound in a color antitriplet state, which then
in turn binds with the remaining quark to form a color singlet. For 
$T \simeq 0$, extensive studies have addressed the formation of QCD 
Cooper pairs; their condensation would then lead to a color superconductor 
\cite{super}. In contrast to 
electromagnetic superconductors, where only the global background 
field (phonons) of the medium can result in a binding of electrons into Cooper 
pairs, we have in QCD a local $qq$ color anti-triplet attraction, 
provided by gluon exchange. Hence diquarks could here exist as local 
``particle'' bound states and not just as momentum state pairs of undefined 
spatial separation. The thermodynamics of a medium of such ``localizable'' 
diquarks was in fact considered some time ago \cite{Ekelin,Sateesh,Anselmino}
\footnote{It should be noted, however, that possible observable
consequences (H-dibaryon, diquark contributions to hadronic structure 
functions from deep inelastic scattering) in the low baryon density region
have not been found.}
Moreover, heavy quark studies \cite{Doering} indicate that the interquark 
potential for a color anti-triplet $QQ$ pair in a deconfined 
medium is attractive and essentially identical to that for a color 
singlet $\Q$ pair. 

\medskip

To reach baryon densities beyond the percolation limit, the nucleons 
evidently have to break up into constituents which are subject to less
or no hard-core repulsion, and a break-up into constituent quarks and 
bosonic diquarks as their excited states fills that requirement.
Since at low $T$ there is little or no thermal agitation, both
could survive for some range of temperature 
and baryochemical potential. 
The resulting picture of the new medium thus parallels somewhat  
that of hadronic matter at $\mu \simeq 0$, where resonance interactions 
lead to a gas of basic hadrons (pions and nucleons) plus their resonance 
excitations. Here we have instead a gas of basic constituent quarks, 
together with the diquark excitations formed through their 
interaction\footnote{We restrict ourselves here to diquarks. In principle,
the formation of multi-quark clusters seems also conceivable.}. The essential 
difference is that the basic ``particles'' now are massive colored fermions,
which can exist only in the colored background field provided by a
sufficiently dense strongly interacting medium.

\medskip

Let us elaborate this structural similarity a little and look at
the possible limits of the two different states of matter with
broken chiral symmetry, hadron gas and constituent quark gas.
In the case of hadronic matter at low $\mu$, we had obtained a
limit through percolation, based on the geometric size of 
hadrons and their overlap \cite{Pomeranchuk}. Such a limit will
presumably exist as well for constituent quarks and diquarks of
comparable size. For constituent quarks, overlapping gluon clouds prevent
an unambiguous definition, and for diquarks, the presence of
equally close alternative binding partners has the same effect.  
We thus expect a continuation of the percolation line into the
deconfined state at larger $\mu$, defining here as well the limit
of chiral symmetry breaking. 

\medskip

An alternative approach to limiting hadronic matter \cite{Hagedorn,DRM} 
is based on the resonance structure of hadronic interactions.   
It leads to an exponentially increasing resonance degeneracy, which
in turn produces singular thermodynamic behavior. The basis for 
this is the possibility of replacing an interacting hadron gas
by an ideal gas of all possible resonances \cite{Beth,DMB},
provided that the interaction is specified by resonance formation. 
The limiting temperature is in this case determined by the range
of the interaction \cite{H-W,HS-Fort}. If the form of the interaction
in a color antitriplet state were similar to that in a color singlet
state, we would expect diquarks to exhibit a corresponding resonance 
pattern. The resulting thermodynamics would then lead to a limiting
temperature for the quark/diquark resonance gas, just as it did for the 
hadronic resonance gas\footnote{The limiting
temperature of the hadronic resonance gas is determined by the
range of the interaction and persists even in the chiral limit of
massless pions. In the original Hagedorn formulation \cite{Hagedorn},
the estimate $T_H \sim m_{\pi}$ arose by assuming the pion mass to
determine the interaction range.}. We note that in both cases, the
geometric argumentation appears as the more general one, since even
a gas containing only extended ground state hadrons leads through
percolation to a limit \cite{Pomeranchuk}. 

\medskip

We thus speculate that the chiral symmetry breaking limit for the
constituent quark state is in nature very similar to that of the hadronic
state, reaching this conclusion either in terms of the percolation of 
spatially extended colored or colorless constituents, respectively, or 
through a conjectured resonance interaction pattern for diquark binding 
as well as for the hadron spectrum.

\bigskip

{\bf 5.\ Quark Plasma vs.\ Quarkyonic Matter} 

\bigskip

We have argued that nuclear matter, with color confinement and chiral
symmetry breaking, is separated from the canonical plasma of massless
quarks (plus some gluons and antiquarks at $T \not= 0$) by an intermediate
phase of massive quarks as basic constituents. Recent arguments dealing
with strongly interacting matter in the large $N_c$ limit have introduced
``quarkyonic'' matter as an intermediate phase. Let us try to understand
how such a state could be related to the massive quark plasma.
We must keep in mind, however, that as long as corresponding features 
of quarkyonic matter are still under discussion, any comparison will
remain tentative. 

\medskip

Let us first consider the relevant color degrees of freedom,
assuming QCD with $SU(N_c)$ as underlying theory.
In a hadron gas, with color singlet constituents, we have only 
one color state, $d^c_{\rm eff} = 1$. In the constituent quark
gas at large $\mu$, there are massive colored quarks, as well as some
massive colored antiquarks for $T>0$, but no free gluons; hence here we 
have $d^c_{\rm eff} \sim N_c$. Finally, in the quark-gluon
plasma, we have $d^c_{\rm eff} = 2N_c + (N_c^2-1)$, with the
first term counting quarks and antiquarks, the second gluons.
For large $N_c$, the three states of matter thus behave
as $1:N_c:N_c^2$, respectively, and in this aspect our picture
results in a behavior similar to that found in the quarkyonic matter
approach. 

\medskip

\begin{figure}[htb]
\centerline{\psfig{file=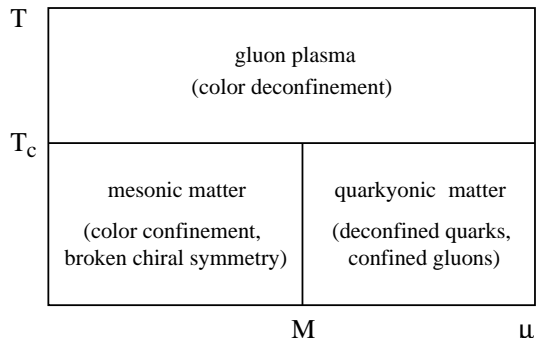,width=7cm}}
\caption{Phase structure of strongly interacting matter for large $N_c$}
\label{quarkyonic}
\end{figure}

However, the limit of large $N_c$ results in several rather drastic
modifications of the phase diagram. The baryon density in the hadronic
regime is given by
\be
n_B \sim \exp\{(\mu - M)/T\},
\ee 
where $M$ again denotes the nucleon mass. Since both $M$ and $\mu$ are
linear in $N_c$, the nuclear matter region is in the large $N_c$ limit
contracted to $\mu=M$. The hadronic regime thus becomes purely
mesonic. Similarly, the quark-gluon plasma, with
\be
T\ln Z(T,\mu_q, V)= (N_c^2-1) \left({\pi^2VT^4\over 45}\right)+
N_f N_c \left({V\over 6}\right) \left[ \left({7 \pi^2T^4\over 30}
\right) + \mu_q^2T^2 + \left( {\mu^4_q\over
2 \pi^2} \right) \right],
\ee
becomes for large $N_c$ gluon dominated for all finite $\mu$ and
$T\geq T_c$. We thus obtain a phase diagram featuring mesonic
matter for $T \leq T_c,~ \mu < M$, and a gluon plasma for
$T\geq T_c$ and all $\mu$. The remaining section, with
$T \leq T_c$ and $\mu \geq M$, is the regime of the proposed
quarkyonic matter. Its effective color degrees of freedom are 
$d^c_{\rm eff} \sim N_c$ and it has non-vanishing baryon density;
it must thus consist of deconfined quarks and confined
gluons, either as quark dressing or as glueballs. In view of
these alternatives, the question of chiral symmetry restoration
remains open. The resulting large $N_c$ phase diagram is illustrated
schematically in Fig.\ \ref{quarkyonic}.

\medskip

\medskip

Our phase diagram thus has several features in common with that
obtained for large $N_c$. An essential difference is that we assume
the onset of chiral symmetry restoration in the quark plasma phase,
at the percolation limit of the extensive massive quarks. With the
quark density given by (see eq.\ (\ref{q-dens}))
\be
n_q = {2 N_c \over 3\pi^2} (\mu^2_q - M_q^2)^{3/2},
\ee  
the percolation condition $n_q=1.2/V_q$ contracts the $\mu$ range
for $N_c \to \infty$ to the point $\mu_q = M_q$, or equivalently
to $\mu = M$. In the limit of large $N_c$, our massive quark phase  
thus shares the fate of nuclear matter: everything is shrunk into
one point, $\mu=M$.

\bigskip

{\bf 6.\ Conclusions} 

\bigskip

Spontaneous chiral symmetry breaking gives the quarks in hadrons
an effective ``constituent'' mass. We have argued that restoration
can occur in two ways,
\begin{itemize}
\item{by an evaporation of the gluonic quark dressing in a hot
environment, or}
\item{by quark percolation in a cold environment, leading to cluster 
fusion of the gluon clouds making up the effective quark mass.}
\end{itemize} 
We suggest that the coincidence of color deconfinement and
chiral symmetry restoration at vanishing baryon density arises
through the first of these two mechanisms.
For low temperature and large baryochemical potential, an evaporation
of the gluon dressing does not appear likely, and hence the effective
quark mass will survive quark deconfinement. We thus obtain a three-phase
structure of matter in QCD (apart from a possible color superconductor),
consisting of
\begin{itemize}
\item{hadronic matter, with the hadronic size as intrinsic scale,
confined quarks and gluons and broken chiral symmetry,}
\item{a plasma of deconfined massive quarks, with their mass
($\sim \langle {\bar\psi}\psi(r_0) \rangle^{1/3}$) as intrinsic scale, 
confined gluons, and broken chiral symmetry,}
\item{a plasma of massless quarks and gluons, with no intrinsic scale,
full color deconfinement and chiral symmetry restoration.}
\end{itemize}
The phase of massive quarks is limited in temperature by evaporation, 
i.e., also by the first of the above alternatives, and in baryon density 
by the second, i.e., by quark percolation. In addition to basic massive
quarks, the quark plasma  can also contain diquarks as local 
quark-quark bound states in the colored background.

\bigskip

{\bf Acknowledgements} 

\bigskip

It is a pleasure to acknowledge helpful and stimulating discussions
and/or correspondence with O.\ Kaczmarek, F.\ Karsch, L.\ McLerran 
and K.\ Redlich.

\end{document}